\documentclass[11pt]{article}

\pdfoutput=1

\usepackage[textwidth=15.5cm,textheight=22.5cm]{geometry}
\usepackage{bm}
\usepackage{amsmath,amssymb,color,graphicx,amscd,amsfonts}
\usepackage{latexsym}
\usepackage{graphicx}
\usepackage{cite}
\usepackage{subfig}
\usepackage{bbm}
\usepackage{ulem}
\usepackage{datetime}

\usepackage{hyperref}

\begin{document}

\thispagestyle{empty}

\setcounter{page}{0}

\begin{flushright} 
SU-ITP-16/14\\
YITP-16-95\\
\yyyymmdddate\today, \currenttime\\[0.2cm]

\end{flushright} 

\vspace{0.1cm}

\begin{center}
{\LARGE

A proposal of the gauge theory description of\\ 
the small Schwarzschild black hole in AdS$_5\times$S$^5$

\rule{0pt}{20pt}  }
\end{center}

\vspace*{0.2cm}

\begin{center}

Masanori H{\sc anada}$^{abc}$
and 
Jonathan M{\sc altz}$^{da}$

\vspace{0.3cm}

$^a${\it Stanford Institute for Theoretical Physics,
Stanford University, Stanford, CA 94305, USA}

$^b${\it Yukawa Institute for Theoretical Physics, Kyoto University,\\
Kitashirakawa Oiwakecho, Sakyo-ku, Kyoto 606-8502, Japan}	

$^c${\it The Hakubi Center for Advanced Research, Kyoto University,\\
Yoshida Ushinomiyacho, Sakyo-ku, Kyoto 606-8501, Japan}

 $^d$
{\it Berkeley Center for Theoretical Physics, University of California at Berkeley,\\
 Berkeley, CA 94720, USA}

\vspace{0.3cm}

hanada@yukawa.kyoto-u.ac.jp

jdmaltz@berkeley.edu or jdmaltz@alumni.stanford.edu

\end{center}

\vspace{1cm}

\begin{abstract}

Based on 4d ${\cal N}=4$ SYM on $\mathbb{R}^{1}\times$S$^3$, a gauge theory description of a small black hole in AdS$_5\times$S$^5$ is proposed. 
The change of the number of dynamical degrees of freedom associated with the emission of the scalar fields' eigenvalues 
plays a crucial role in this description. 
By analyzing the microcanonical ensemble, the Hagedorn behavior of long strings at low energy is obtained. Modulo an assumption based on the AdS/CFT duality for a large black hole,  
the energy of the small ten-dimensional Schwarzschild black hole $E\sim 1/(G_{\rm 10,N}T^7)$ is derived. 
A heuristic gauge theory argument supporting this assumption is also given.
The same argument applied to the ABJM theory correctly reproduces the relation for the eleven-dimensional Schwarzschild black hole. 
One of the consequences of our proposal is that the small and large black holes are very similar when 
seen from the gauge theory point of view.

\end{abstract}


\newpage

\section{Introduction}

\begin{figure}[htbp]
\begin{center}
\rotatebox{0}{
\scalebox{0.6}{
\includegraphics{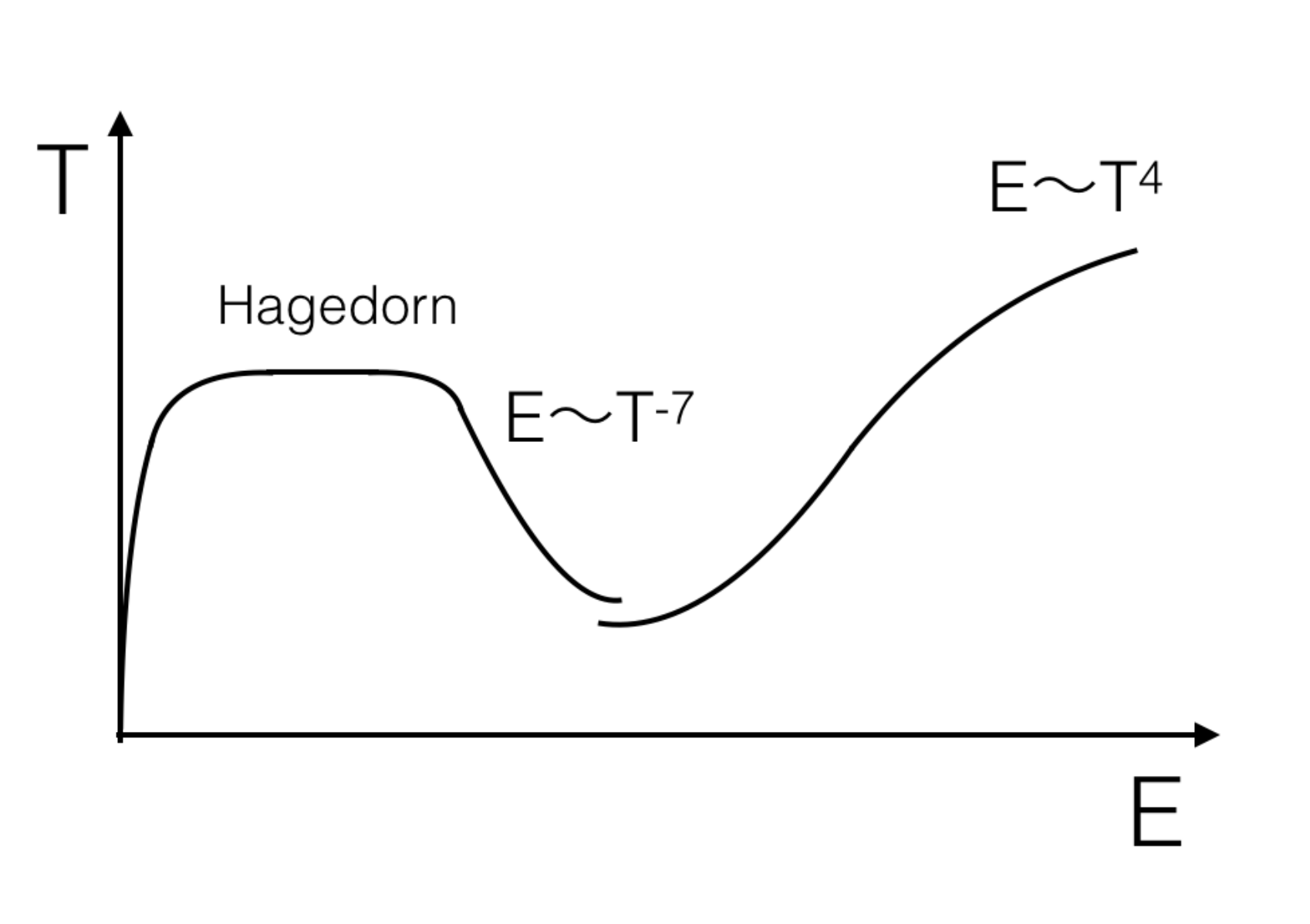}}}
\end{center}
\caption{
The microcanonical $E$-vs-$T$ phase diagram of  4d SYM on $\mathbb{R}^{1}\times$S$^{3}$ at strong coupling ($\lambda=g_{\rm YM}^2N\gg 1$), obtained 
{\it by assuming the validity of the AdS/CFT duality}.}\label{phase_diagram}
\end{figure}

\hspace{0.25in}Gauge/gravity duality \cite{Maldacena:1997re} is believed to be a key idea in resolving the black hole information paradox. 
Witten \cite{Witten:1998qj} conjectured that 4d ${\cal N}=4$ super Yang-Mills (SYM) theory compactified on a three-sphere S$^3$, 
whose action is given by
\begin{eqnarray}
S=\frac{1}{g_{\rm YM}^2}\int d^4x\ {\rm Tr}\left(
\frac{1}{4}F_{\mu\nu}^2+\frac{1}{2}(D_\mu X_M)^2+\frac{1}{4}[X_M,X_{M'}]^2
-\frac{1}{2}X_M^2
+
(fermion)
\right), 
\end{eqnarray}
can describe a black hole (BH) in AdS$_5\times$S$^5$. 
Here the gauge group is SU($N$) and $X_M$ ($M=1,\cdots,6$) are $N\times N$ Hermitian matrices. 
We consider 't Hooft large-$N$ limit, $g_{\rm YM}^2\propto N^{-1}$, and the radius of the S$^{3}$ is set to 1. 
If the conjecture is correct, then the dual gravity description suggests the following behavior in phase diagram of the microcanonical ensemble 
at strong coupling (see e.g. 
Sec.~3.4.1 of \cite{Aharony:1999ti}):
\begin{itemize}
\item
When the energy $E$ is large enough, a large AdS-BH, which fills the S$^5$ direction, is formed. 
The energy scales as $E\sim\frac{R_{\rm AdS}^{11}T^4}{G_{\rm 10,N}}$ at high temperature $T$, 
where $G_{\rm 10,N}$ and $R_{\rm AdS}$ are the ten-dimensional Newton constant and the AdS radius, respectively. 
Note that the large AdS-BH has a positive specific heat. 

\item
The large AdS-BH shrinks as the energy is decreased and the temperature goes down.  
When the Schwarzschild radius becomes of order $R_{\rm AdS}$, the BH localizes along the S$^5$ 
and can be regarded as a ten-dimensional BH. This transition is of first order \cite{Dias:2016eto}, 
and the BH becomes hotter after the localization.\footnote{
The authors would like to thank Jorge Santos for the clarification.}
When the Schwarzschild radius becomes much smaller than $R_{\rm AdS}$, the BH should behave
like the ten-dimensional Schwarzschild BH in flat spacetime, $E\sim\frac{1}{G_{\rm 10,N}T^7}$. We will call this localized BH, the small BH. 
Note that the small BH has a negative specific heat. 

\item
As the small BH shrinks towards the string scale, the description of it as a bunch of long strings become better. 
The system shows the famous Hagedorn behavior, $E\propto S\propto L$, where $S$ is the entropy 
and $L$ is the length of the strings. 

\item
Finally, when the energy is very small, the system is well described as a gas of short strings. 

\end{itemize}
The relation between the energy and temperature is shown in Fig.~\ref{phase_diagram}. 

Although the gauge/gravity duality conjecture has not been proven, there is accumulating evidence that it is most likely correct. 
Hence the majority of string theorists believe that 4d SYM has the same phase diagram.  
It is crucial however to understand this phase diagram directly from the gauge theory for several reasons.
First of all, the duality has been poorly tested at finite temperature. As far as we know, the only quantitative tests currently understood\cite{Berkowitz:2016jlq}\footnote{
Analytic approaches for deriving nontrivial temperature dependence have been discussed in \cite{Smilga:2008bt,Wiseman:2013cda}. 
}, are for 
the duality \cite{Itzhaki:1998dd} between the type IIA black zero-brane and D0-brane quantum mechanics \cite{Witten:1995im,Banks:1996vh,deWit:1988ig}, which is analogous to the large AdS-BH in 4d ${\cal N}=4$ SYM. 
Hence we need to test gauge/gravity duality for a small BH. 
Furthermore, if we use the duality to understand the quantum gravitational aspects of black holes, we have to solve gauge theory. 
If we assumed the validity of the dual gravity description and used the gravitational description to explain gravity, 
we would be just be assuming the answer to answer the question.  

Previously, the gauge theory description of the large AdS-BH, the Hagedorn parameter region, and string gas parameter region have been understood at least qualitatively 
(see e.g. \cite{Aharony:1999ti}).  
In this paper, we propose a simple gauge theory description of the small BH. 
By assuming the validity of the AdS/CFT correspondence {\it for the large black hole},
we derive the relation between the energy and temperature {\it of the small black hole}, $E\sim 1/(G_{\rm 10,N}T^7)$, at strong coupling. 
We also give a heuristic explanation supporting this assumption based only on gauge theory. 
In addition, we will show that the same picture correctly reproduces the Hagedorn behavior. 
In short, 
\begin{itemize}
\item
The large black hole is described by a bound state of all the eigenvalues of scalar fields $X_M$. 
All $N^2$ matrix entries are excited. 

\item
Suppose some of the eigenvalues are emitted, after which and only $N_{\rm BH}< N$ eigenvalues form a bound state.  
Such matrices describe the small black hole. The black hole is smaller when $N_{\rm BH}$ is smaller.\footnote{
The idea that the size of the matrix blocks changes with the energy has also been an important ingredient of a proposal for a description 
of the Schwarzschild black hole in the Matrix Model of M-theory \cite{Banks:1997hz}. 
} 

\end{itemize}
%

This paper is organized as follows. 
In Sec.~\ref{sec:microscopic_picture}, we remind the readers how the microscopic, stringy degrees of freedom can be read off from the fields (matrices) in 4d SYM. Two seemingly different, but actually equivalent, pictures -- `open strings+D-branes' and closed strings -- are introduced, 
and the meaning of the emission of the D-branes (eigenvalues) from the BH \cite{Berkowitz:2016znt,Berkowitz:2016muc} is explained. 
Sec.~\ref{sec:main} is the main part of this paper. We propose a gauge theory description of the small black hole, 
and obtain the relationship between the energy, entropy and temperature expected from the conjectured gravity dual 
modulo a technical assumption explained at the end of the section. 
In Sec.~\ref{sec:others}, we suggest that the same picture can hold for a rather generic class of theories 
holographically dual to superstring/M-theory. We study the ABJM theory as an example
and derive the right energy-temperature relation of the 11d Schwarzschild black hole.  
\begin{center}
\section*{Note Added}
\end{center}
\hspace{0.25in}While this work has been in progress, we have learned that Leonard Susskind had essentially the same idea independently. 
He conjectured the small black hole is described by a small sub-matrix, and considered a possibility of 
making the `box' (AdS space) smaller in order to remove the degrees of freedom which are not needed for describing the small BH. 
In terms of gauge theory, this means a truncation to U$(N_{\rm BH})$. Then he assumed the `corresponding principle' 
which relates the large and small black holes. On the gauge theory side, mathematically, this is exactly what we have done 
in order to derive $E\sim 1/(G_{10,{\rm N}}T^7)$.  
We would like to thank him for stimulating discussions and collaboration toward the end of the project. \\
\section{Stringy interpretation of the field theory degrees of freedom}\label{sec:microscopic_picture}
\hspace{0.25in}In this section, we explain how the stringy micro-states of a black hole are encoded into gauge theory. 
There are two seemingly different pictures, (1) the bound states of D-branes and strings \cite{Witten:1995im} 
and (2) long, winding strings \cite{Susskind:1993ws,Hanada:2014noa}. Here we explain how they are related to each other.  
 
Firstly let us consider the D0-brane quantum mechanics picture\cite{Witten:1995im,Banks:1996vh,deWit:1988ig}, the generalization to generic gauge theories including 4d ${\cal N}=4$ SYM is straightforward.
We work in the Hamiltonian formulation \cite{Kogut:1974ag}. 
The gauge field is set to zero, $A_t=0$, and the physical Hilbert space is obtained 
by acting traces of products of scalars $X_M(M=1,2,\cdots,9)$ on the vacuum state.\footnote{
More precisely speaking, the adjoint fermions exist as well.
The gauge-singlet condition follows from the Gauss-law constraint. 
}    
When we follow the usual D-brane effective theory point of view \cite{Witten:1995im}, the diagonal components $X_{M,ii}$ are regarded as 
the position of $i$-th D0-brane in ${\mathbb R}^9$, and the off-diagonal components $X_{M,ij}$ describe 
open strings connecting $i$-th and $j$-th D0-branes. Note that these strings are oriented. 
When $X_{M,ij}$ is large, a lot of strings are excited between $i$-th and $j$-th D0-branes. 
This is picture (1). 

\begin{figure}[htbp]
\begin{center}
\rotatebox{0}{
\scalebox{0.3}{
\includegraphics{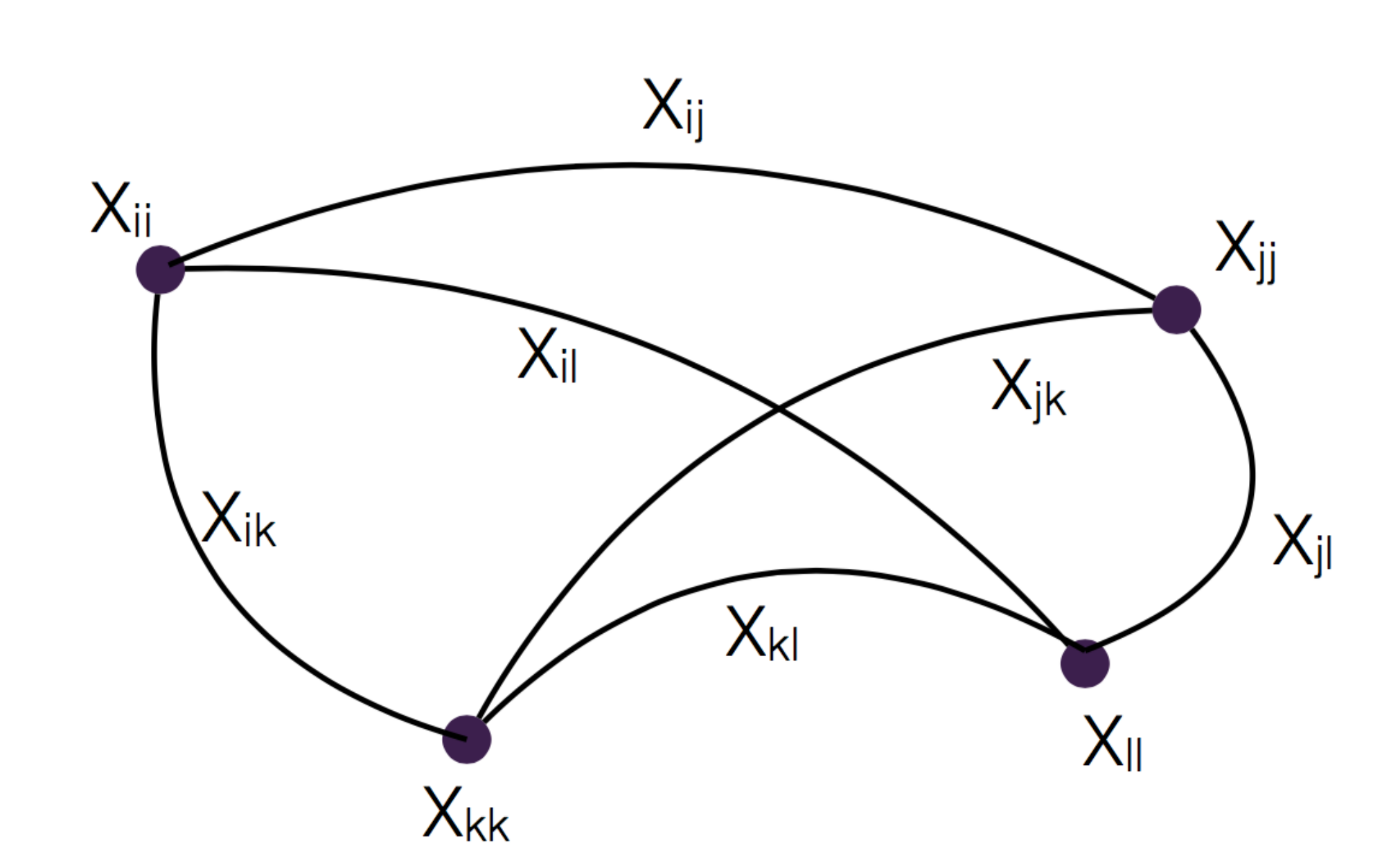}}}
\end{center}
\caption{Interpretation of matrix components as sites and links of a nonlocal lattice.}\label{figure_lattice}
\end{figure}

\begin{figure}[htbp]
\begin{center}
\rotatebox{0}{
\scalebox{0.3}{
\includegraphics{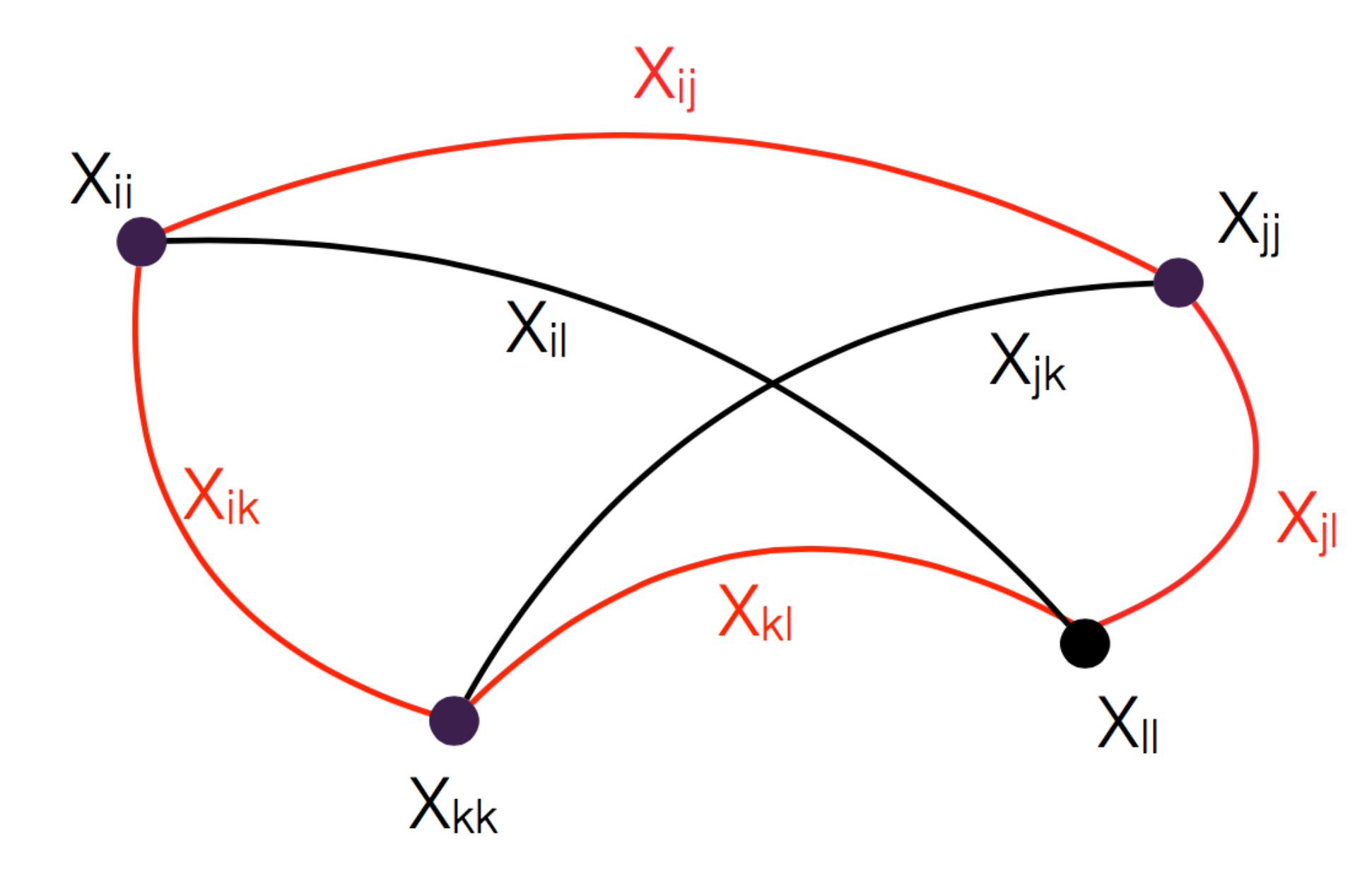}}}
\end{center}
\caption{$X_{M_1,ik}X_{M_2,kl}X_{M_3,lj}X_{M_4,ji}$ in ${\rm Tr}(X_{M_1}X_{M_2}X_{M_3}X_{M_4})|{\rm \it{Vac}}\rangle$.}\label{figure_loop_1}
\end{figure}

\begin{figure}[htbp]
\begin{center}
\rotatebox{0}{
\scalebox{0.3}{
\includegraphics{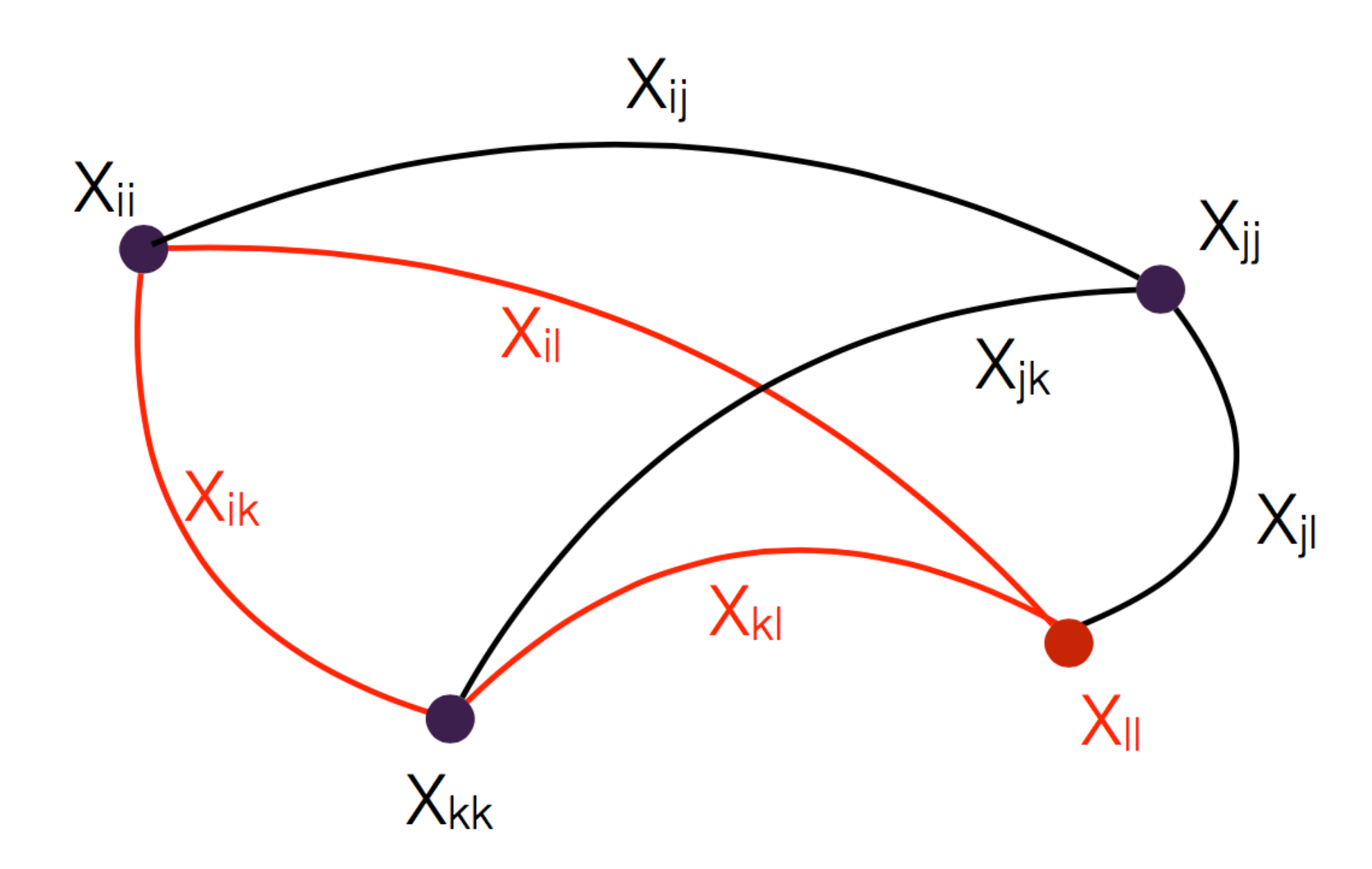}}}
\end{center}
\caption{$X_{M_1,ik}X_{M_2,kl}X_{M_3,ll}X_{M_4,li}$ in ${\rm Tr}(X_{M_1}X_{M_2}X_{M_3}X_{M_4})|{\rm \it{Vac}}\rangle$.}\label{figure_loop_2}
\end{figure}

In order to go to picture (2), let us regard the D0-branes and open strings as sites and links of a non-local lattice 
(Fig.~\ref{figure_lattice}). Here we use the adjective ``nonlocal" because 
all pairs of sites can be directly connected by links.  
Gauge-invariant states are made of closed loops.\footnote{
This idea is not new. See e.g. \cite{Asplund:2008xd}. 
}
For example, if we consider 
${\rm Tr}(X_{M_1}X_{M_2}X_{M_3}X_{M_4})|{\rm \it{Vac}}\rangle$; 
$X_{M_1,ik}X_{M_2,kl}X_{M_3,lj}X_{M_4,ji}$ with different $i,j,k,l$ is a closed loop 
made of four links (Fig.~\ref{figure_loop_1}), while 
$X_{M_1,ik}X_{M_2,kl}X_{M_3,ll}X_{M_4,li}$ with different $i,k,l$ is a closed loop 
made of three links and one site (Fig.~\ref{figure_loop_2}). 
Hence the black hole, which is a bound state of D-branes and open strings, is naturally regarded as a long, winding string in the lattice description. 
(More precisely, a few long strings.) 
The maximum possible number of string bits (open strings) is the maximum possible length of a single trace operator, 
which is of order $N^2$. This upper bound appears because single trace operators beyond this bound can be expressed by using 
shorter operators. 
Note that this is slightly different from the `correspondence principle' \cite{Susskind:1993ws,Horowitz:1996nw} in the usual sense, 
that the black hole is always described by strings in this picture \cite{Susskind:1993ws,Hanada:2014noa}. 

When the number of spatial dimension is nonzero, the gauge fields form other links; the picture should be clear if one imagines a lattice discretization of space-time. 
Then the link variables also become strings; generic long string states are expressed by Wilson loops (QCD strings) with scalar insertions. 
Here the black brane can be regarded as a condensation of long strings. For more details, see \cite{Hanada:2014noa}.\\ 

\subsection{Emission of D-branes and black hole evaporation}

\hspace{0.25in}The dynamics of D-branes play an important role in the matrix (gauge theory) description of the black holes \cite{Banks:1996vh}. 
Suppose the D-branes form well separated and localized bunches 
consisting of $N_1, N_2, \cdots$ D-branes. 
Then strings inside each bunch are short, light and can be very excited, 
while strings connecting different bunches are long, heavy and cannot be excited much. 
In terms of matrices, such a configuration is expressed by almost block-diagonal matrices 
with block sizes $N_1, N_2, \cdots$. 
Each block is regarded as long strings with maximum length $N_1^2, N_2^2, \cdots$, respectively. 
Typically, the $i$-th block carries the energy and entropy of order $N_i^2$. 

Next let us consider the emission of eigenvalues, following \cite{Berkowitz:2016znt,Berkowitz:2016muc}.\footnote{
In the theories with flat directions, the emission is inevitable. In 4d ${\cal N}=4$ SYM on $\mathbb{R}^{1}\times$S$^3$, though there are no flat directions, 
emission still plays an important role, as we will see shortly. 
} 
When one of the D-branes is emitted, open strings between the emitted D-brane and the others become heavy and decouple from the dynamics. 
Hence fully noncommutative $N\times N$ matrices turn to block diagonal forms, 
\begin{eqnarray}
\left(
\begin{array}{cc}
X^M_{\rm BH} &  0\\
0 & x^M
\end{array}
\right), 
\end{eqnarray}
where $X^M_{\rm BH}$ are fully noncommutative $(N-1)\times (N-1)$ matrices and $x^M$ describe the position of emitted D-brane\footnote{
Here we implicitly took a gauge in which the emitted D-brane is described by the $(N,N)$-components. 
Technical details about this gauge choice will be explained in Sec.~\ref{sec:gauge_fixing}. 
}. This is the Higgs mechanism. 
The number of light physical degrees of freedom decrease from $\sim N^2$ to $\sim (N-1)^2+1$, while the energy is conserved during emission. Hence the energy per degree of freedom ($\simeq$ temperature) increases.  
As the emission continues more eigenvalues can be emitted and the black hole becomes hotter and hotter. 
The negative specific heat of the black hole follows from this simple fact. For details, see \cite{Berkowitz:2016znt,Berkowitz:2016muc}. 

In the next section, we show that the emission of D-branes can explain the properties of the small black hole in AdS$_5\times$S$^5$
described by 4d SYM on $\mathbb{R}^{1}\times$S$^{3}$. 
\section{Analysis of the microcanonical ensemble}\label{sec:main}

\subsection{The large black hole (large energy and $\lambda\gg 1$)}
\begin{figure}[htbp]
\begin{center}
\rotatebox{0}{
\scalebox{0.5}{
\includegraphics{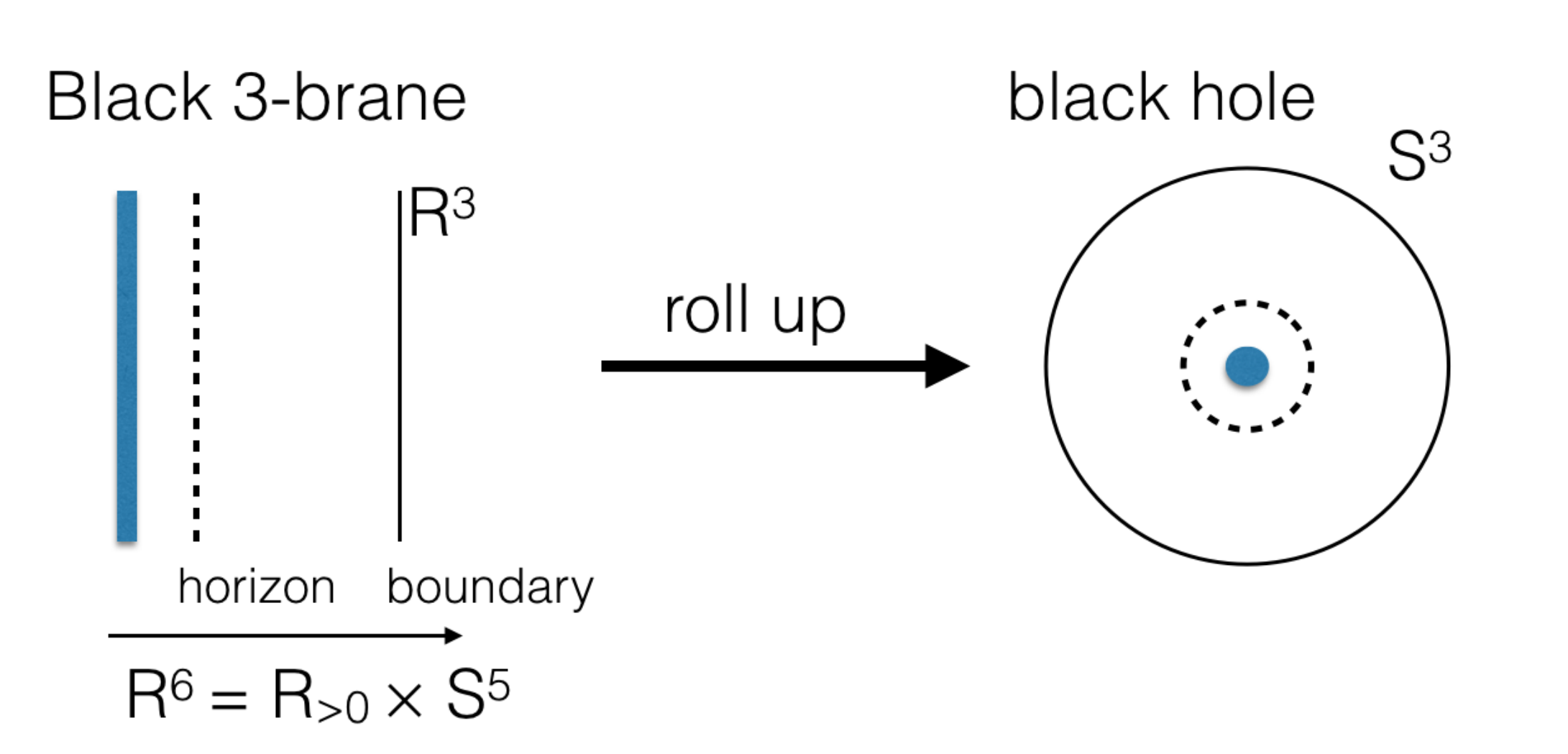}}}
\end{center}
\caption{
AdS black hole is obtained by rolling up black 3-brane.}\label{rolling_up_D3}
\end{figure}

\hspace{0.25in}The large AdS$_5$ black  hole is obtained by rolling up a black three-brane with charge $N$. 
The black brane is a bound state of all $N$ eigenvalues (D3-branes);  as is the large black hole. See Fig.~\ref{rolling_up_D3}. 
As we declared in the introduction, we assume the AdS/CFT duality between SYM and the large BH is correct. 
In principle (and within a ten-year span, probably in practice), the duality can be tested by Monte Carlo simulation. 
For recent numerical studies, see e.g. \cite{Catterall:2012yq}. 

\begin{figure}[htbp]
\begin{center}
\rotatebox{0}{
\scalebox{0.4}{
\includegraphics{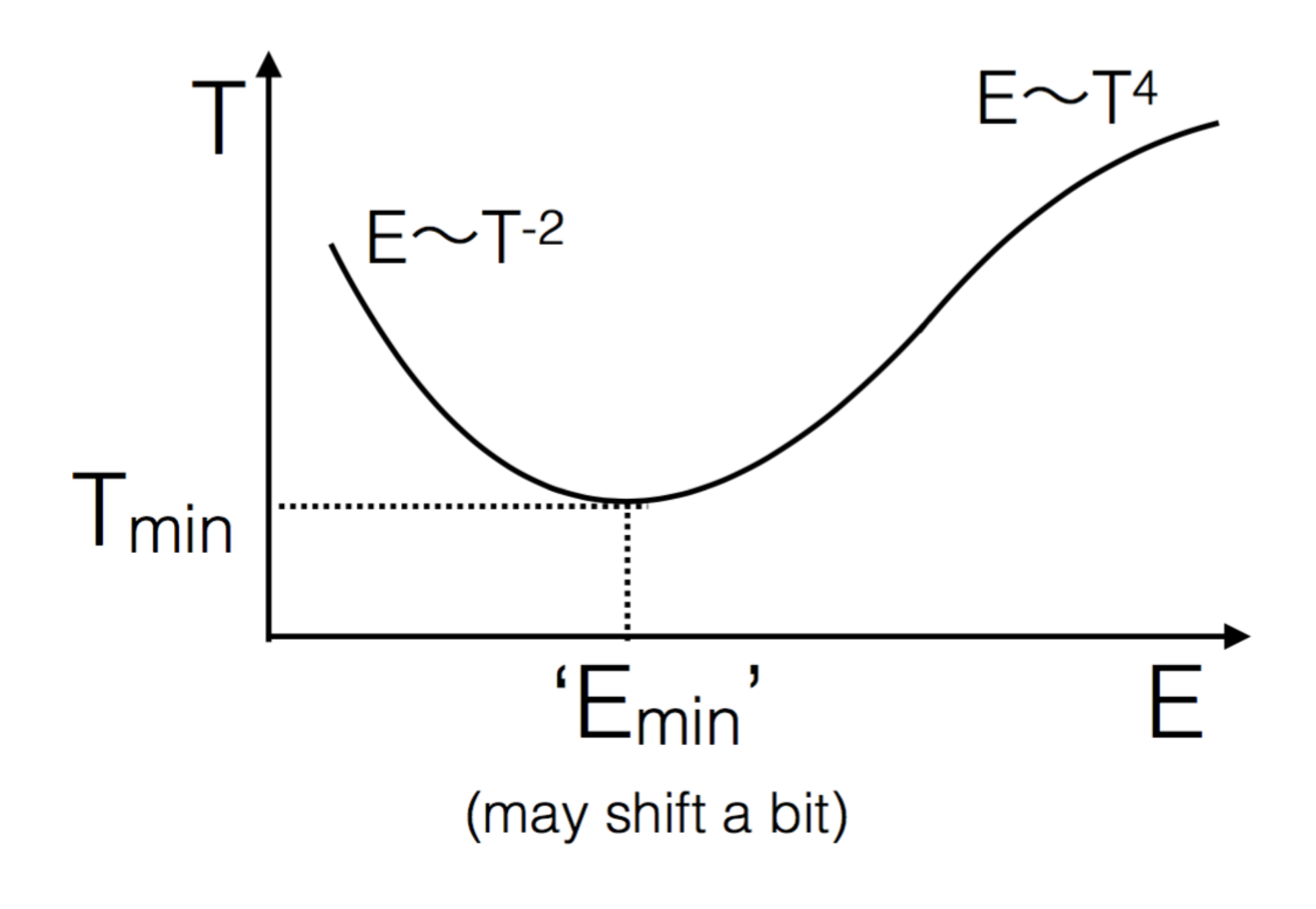}}}
\end{center}
\caption{The microcanonical $E$-vs-$T$ phase diagram of AdS$_5$-BH which is fills the S$^5$ completely. $E<E_{\rm min}$ does not have a counterpart in 4d SYM; it is unstable with respect to the Gregory-Laflamme 
instability along the S$^5$.}\label{AdS5-BH-2}
\end{figure}

When the black hole on the gravity side fills the S$^5$ completely, the AdS$_5$ BH should be used in the dual gravity calculation. 
It has the minimum temperature $T_{\rm min}$; see Fig.~\ref{AdS5-BH-2}. 
Let us call the energy at $T=T_{\rm min}$ to be $E_{\rm min}$, and the solutions at $E>E_{\rm min}$ and $E<E_{\rm min}$ to be 
`large' and `small' AdS$_5$ black holes, respectively. (Hence $E_{\rm min}$ is the minimum energy of the `large' BH.) To avoid confusion we will use small BH to refer to a 10d Schwarzschild black hole and small AdS$_5$ BH to refer to a small AdS black hole which still fills the S$^5$.

The small AdS$_5$ BH solution at $E<E_{\rm min}$ 
does not have a counterpart in 4d SYM; it is unstable with respect to the Gregory-Laflamme instability \cite{Gregory:1993vy} along the S$^5$, 
and hence the 10d BH becomes the appropriate description.  
From the point of view of gauge theory, it can be understood as follows. 
In order for a bound state of eigenvalues (non-commutative block) to be formed, 
(almost) all of the off-diagonal elements must be excited, which costs a lot of energy. 
Hence the AdS$_5$ BH black hole can exist only when the energy is large enough. 

Strictly speaking, 
the `minimum energy' $E_{\rm min}$ corresponding to the instability
might be slightly different from the energy at $T_{\rm min}$; 
they should be of the same order but it is hard to determine the order one factor. 
A dual gravity analysis such as \cite{Dias:2016eto} may provide us with a concrete number.
In the following we will not consider order one coefficients which are sensitive to this ambiguity. 

\subsection{Emission of eigenvalues}
\hspace{0.25in}
The above argument, however, does not take into account the emission of eigenvalues.\footnote{
The speculation that a subset of D-branes describes the small black hole existed for quite some time; see e.g. \cite{Asplund:2008xd}. 
}
Even if the energy is not large enough to bind all $N$ eigenvalues, it may still be possible to bind $N_{\rm BH}$ eigenvalues, with $N_{\rm BH}<N$. 
Note that, because the space is compactified, there is no superselection of vacua; 
the value of the scalar field $X^M (M=1,2,\cdots,6)$ should be determined dynamically, 
like in the Matrix Model of M-theory \cite{Banks:1996vh}. 
Also note that, unlike the Matrix Model of M-theory, this theory does not possess flat directions, 
because the scalars have a mass proportional to the inverse of the S$^3$ radius. Hence the emitted particles do not roll to infinity; 
they form a finite density gas and that can be absorbed again by the black hole. At some point, the emission and the absorption rates can balance. 

\begin{figure}[htbp]
\begin{center}
\rotatebox{0}{
\scalebox{0.2}{
\includegraphics{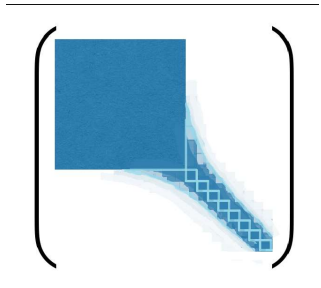}}}
\end{center}
\caption{A more precise representation of the shape of the matrices describing a small black hole. 
In addition to the elements shown in \eqref{matrix_small_BH}, there is some `fuzziness' which 
describes short open strings stretched between nearby D-branes. The light blue lines indicate where the blocks would be if the matrix were block diagonal. The color refers to the magnitude of the matrix elements, darker blue implying a larger magnitude.
}\label{fig:actual_shape}
\end{figure}

We consider a mixed state of BH and gas in 4d ${\cal N}=4$ on $\mathbb{R}^1\times$S$^3$. 
Our proposal is that a black hole consisting of $N_{\rm BH}$ eigenvalues 
and a gas consisting of $N_{\rm gas}$ particles, where $N=N_{\rm BH}+N_{\rm gas}$, 
can be described by matrices of the form
\footnote{
Here we implicitly took a gauge in which the `black hole' comes to the  upper-left corner. 
In Sec.~\ref{sec:gauge_fixing} we show how this gauge choice can be achieved, and that the gauge fixing and Faddeev-Popov terms 
are negligible.
Strictly speaking, in addition to the elements shown in \eqref{matrix_small_BH}, there is some `fuzziness' which 
describes short open strings stretched between nearby D-branes; see Fig.~\ref{fig:actual_shape}. 
Such a correction is negligible in the situations we consider below, where both $N_{\rm BH}/N$ is small but of order $N^0$.   
} 
\begin{eqnarray}
\left(
\begin{array}{c|cccc}
X^M_{\rm BH} & & & & \\
\hline
& x^M_1 & & & \\
& & x^M_2 & & \\
& & & \ddots & \\ 
& & & & x^M_{N_{\rm gas}}
\end{array}
\right), 
\label{matrix_small_BH}
\end{eqnarray}
where $X_{\rm BH}$ is an $N_{\rm BH}\times N_{\rm BH}$ matrix.

Intuitively, 
\begin{itemize}
\item
In order for a bound state of eigenvalues (non-commutative block) to be formed, 
(almost) all off-diagonal elements must be excited, which costs a lot of energy. 

\item
If the total energy of the system is big enough, say $E\sim N^2$, 
(almost) all off-diagonal elements can be excited. Hence $N=N_{\rm BH}$, $N_{\rm gas}=0$ can be realized. 

\item
If the energy is not that big, then $N_{\rm BH}$ becomes smaller. 
The number  of degrees of freedom decreases dynamically, and the temperature of the system goes up. 
At some point, (if the total energy is not too small) $N_{\rm BH}$ can become small enough 
so that off-diagonal elements in $X_{\rm BH}$ (number of degrees of freedom $\sim N_{\rm BH}^2$) 
can be excited. 

\item
In the closed string picture: the string wants to become as long and winding as possible with the available amount of energy. It does this in order to gain more entropy. 
When the energy is not big enough, i.e. $N_{\rm BH}$ becomes too big, it ends up with a long but not very wound string. 
Then $N_{\rm BH}$ becomes smaller so that the string can take more complicated shapes. 

\end{itemize}

Strictly speaking, there are strings connecting different gas-branes, and also the ones connecting the BH and gas-branes. See Fig.~\ref{fig:actual_shape}.
Here we are assuming that most of them are long, heavy and do not play an important role. 

In our calculation, we neglect the interaction between the emitted eigenvalues and the small BH, 
and treat it similarly as the `large BH' in the truncated U$(N_{\rm BH})$ gauge theory.

\subsubsection{Gauge fixing}\label{sec:gauge_fixing}
\hspace{0.25in}
Here we explain how the (almost) block-diagonal form \eqref{matrix_small_BH} and Fig \ref{fig:actual_shape}\, can be obtained. 
Let us first introduce the maximally-diagonal gauge \cite{Azeyanagi:2009zf}. 
This gauge condition is based on the implicit assumption of the `D-brane+open string' picture -- 
diagonal elements are large, off-diagonal elements are small. 
However this is actually a gauge-dependent statement. Hence, we impose the condition that 
`the matrices are as close to simultaneously diagonal as possible'. 
For that purpose we introduce $R_{ij}$ defined by
\begin{eqnarray}
R_{ij}\equiv\int d^3x \sum_{M=1}^6 |X_{M,ij}|^2. 
\end{eqnarray}
Under the gauge transformation $X_M\to\Omega X_M\Omega^{-1}$, $R_{ij}$ transforms as 
\begin{eqnarray}
R_{ij}\to R_{ij}(\Omega)\equiv\int d^3x \sum_{M=1}^6 |(\Omega X_M\Omega^{-1})^{ij}|^2. 
\end{eqnarray}
We choose $\Omega=\Omega_{\rm max}$ which maximizes ${\rm Tr}R = \sum_{i=1}^N R_{ii}(\Omega)$. 
Unless there are accidental degeneracies, such a $\Omega_{\rm max}$ is unique up to U$(1)^{N-1}$ 
and simultaneous permutations of rows and columns. 
The $\tilde{X}_M\equiv\Omega_{\rm max} X_M \Omega_{\rm max}^{-1}$ are `as close to simultaneously diagonal as possible'. 

Now let us apply this gauge choice to the situation under consideration. Among the $N$ D-branes, $N_{\rm BH}$ 
form the small black hole. We assume both $N$ and $N_{\rm BH}$ are parametrically large, 
and $N_{\rm BH}/N$ is small but of order $N^0$. Then, the small black hole is stable, 
namely the value $N_{\rm BH}$ does not fluctuate much. 
The $N_{\rm BH}$ diagonal elements in the maximally diagonal gauge should then form a bunch 
(implying that the off-diagonal elements connecting them are not small), 
while other $N_{\rm gas}=N-N_{\rm BH}$ diagonal elements are spread out. 
We can (partly) fix the ambiguity of the permutation of rows and columns 
by putting the $N_{\rm BH}$ size bunch in the upper-left corner of the matrices. 
Then the matrices takes the form \eqref{matrix_small_BH}. 

In our arguments throughout the paper, intuitively, we are `truncating' (or freezing out the gas degrees of freedom to reduce)
$X$ to $\tilde{X}_{\rm BH}$. 
Note that this `truncation' makes sense only in the microcanonical ensemble, and for a fixed value of the energy; 
when the extra energy is added, the bunch size increases in the original theory, while the bunch size is fixed and temperature increases in the truncated theory.
The gauge fixing and the Faddeev-Popov terms associated with the maximally-diagonal gauge should be taken into account. 
But the maximally-diagonal gauge condition is rather nontrivial (note that $\Omega_{\rm max}$ is time-dependent!) 
and it is difficult to write down the gauge fixing and the Faddeev-Popov terms. 
In the situation we have in mind, however, these terms are negligible up to the interaction with emitted particles. This is
because, at each time $t$ the truncated action written by $\tilde{X}_{\rm BH}(t)$ 
is the same as the original action written by $X(t)$ to the leading order (i.e. $O(N^2)$ parts of the actions agree), 
as long as $N_{\rm BH}/N$ is of order $N^0$. 
As long as we consider physics of the small black hole (the bunch of $N_{\rm BH}$ D-branes), 
any gauge-equivalent profiles give simply the same path-integral weight as the truncated theory, 
to the leading order in $1/N$. 

Another way to fix the gauge is to introduce an `external field'. 
Let $P_n$ be a projector to the $n\times n$ block, $P_n={\rm diag}(1,1,\cdots,1,0,0,\cdots,0)$ with $n$ 1's and $N-n$ 0's. 
We can introduce an `external field' which pushes BH to the upper-left corner, for example as 
\begin{eqnarray}
S_{\rm ext} = c\sum_M{\rm Tr}\left(X_M-P_{N_{\rm BH}} X_M P_{N_{\rm BH}}\right)^2. 
\end{eqnarray}
Note that this term manifestly breaks the full SU$(N)$ symmetry. 
Then the BH is pushed to the upper-left corner by 
minimizing $S_{\rm ext}$ which occurs at positive value $c$ which is small but $O(N^0)$. 
After taking $N\to\infty$, we turn off $c$. This is nothing but the usual prescription used for detecting spontaneous symmetry breaking.

\subsection{10d Schwarzschild at $1/\lambda \ll N_{\rm BH}/N < 1$ ($\lambda\gg 1$)}

\hspace{0.25in}We will now make this scenario more precise. 
The AdS-Schwarzschild BH sitting at the origin of AdS corresponds to a bound state of eigenvalues around $X^M=0$.  

The energy is 
\begin{eqnarray}
E_{\rm total} = E_{\rm BH}  + E_{\rm gas},  
\end{eqnarray}
and roughly speaking, 
\begin{eqnarray}
E_{\rm BH} \sim N_{\rm BH}^2 =O(N^2), 
\qquad
E_{\rm gas} \sim (N-N_{\rm BH}) =O(N). 
\end{eqnarray}
Hence we ignore $E_{\rm gas}$. 

We have to determine $N_{\rm BH}$ for given $E\simeq E_{\rm BH}<E_{\rm min}(N,g_{\rm YM}^2)$. 
Let us firstly give a heuristic gauge theory argument which does not rely on the gravity dual. 
Suppose the action is dominated by the $\frac{N}{\lambda}{\rm Tr}[X_M,X_{M'}]^2$-term, which should be true at strong coupling. 
Note that the coupling $\lambda$ disappears in terms of $X'\equiv \lambda^{-1/4}X$; 
hence the eigenvalues of $X'$ are of order 1, and the eigenvalues of $X$ scale as $\lambda^{1/4}$
 when $g^{2}_{\rm YM}$ is varied.  
When the bunch size decreases to $N_{\rm BH}$, the radius of the bunch scales as $\lambda_{\rm BH}^{1/4}$, assuming the interactions with the emitted branes do not affect the size of the bunch significantly. 
Here $\lambda_{\rm BH}=g_{\rm YM}^2N_{\rm BH}=\frac{N_{BH}}{N}\lambda$,
and we assumed $\lambda_{\rm BH}\gg 1$. 
It would be natural\footnote{
The existence of another scale $R_{{\rm S}^3}=1$ makes the situation subtle. 
When $(\lambda_{\rm BH}/\lambda)^{-1/4}\gg1$, the two energy scales $(\lambda_{\rm BH}/\lambda)^{-1/4}$ and $R_{{\rm S}^3}=1$ are clearly separated. Then energy scale should be dominated by $(\lambda_{\rm BH}/\lambda)^{-1/4}$ which is the size of the eigenvalue distribution. 
} to expect that the typical energy scale is set by the inverse of the eigenvalue distribution, $\lambda_{\rm BH}^{-1/4}$. 
Then, combined with the fact that the large BH should have $T_{\rm min}(N,g_{\rm YM}^2)\sim 1$, we obtain\footnote{
We would like to thank Juan Maldacena and Kostas Skenderis 
for pointing out several miscalculations in the first version of this paper, including this part. 
It helped us debug a few wrong points in the argument. 
} 
\begin{eqnarray}
T_{\rm min}(N_{\rm BH},g_{\rm YM}^2)\sim (\lambda_{\rm BH}/\lambda)^{-1/4}\sim\alpha^{-1/4},
\qquad
\alpha\equiv\frac{N_{\rm BH}}{N},   
\end{eqnarray}
and combined with the 't Hooft counting and $E_{\rm min}(N,g_{\rm YM}^2)\sim N^2$, 
\begin{eqnarray}
E_{\rm min}(N_{\rm BH},g_{\rm YM}^2)\sim N_{\rm BH}^2(\lambda_{\rm BH}/\lambda)^{-1/4}
=N^2\alpha^{7/4}, 
\qquad
S_{\rm min}(N_{\rm BH},g_{\rm YM}^2)\sim N^2_{\rm BH}
=N^2\alpha^2. 
\label{eq:NBH-dependence}
\end{eqnarray} 

A better argument inspired by string theory goes as follows.
In the dual gravity picture, the `smallest large black hole' is a bunch of eigenvalues filling AdS$_5$ almost completely. 
In the D-brane picture, AdS$_5$ is made of ${\mathbb R}^1\times$S$^3$ and 
${\mathbb R}_{>0}$, where ${\mathbb R}_{>0}$ is the radial coordinate 
of the transverse ${\mathbb R}^6$. 
Intuitively, the boundary S$^3$ is almost touching the D-branes.
When the bunch shrinks, the radius becomes smaller by a factor of $(\lambda_{\rm BH}/\lambda)^{1/4}$. 
In order to measure the energy of this bunch, we imagine a sphere right outside of the bunch, 
and consider only the interactions between D-branes and open strings inside this sphere.
Note that this restriction is natural from the gauge theory, or `open string', point of view, in which the gravitational back-reaction is not included. 
From the dual gravity theory point of view (`closed string picture' in the sense of usual open string/closed string duality), 
a naive truncation near the horizon is very problematic. 
As we will see shortly, the counterpart of this restriction on the gravity side is something different; 
the curvature radius changes as well on the gravity side. 
Mapping back to the gauge theory, the energy of this bunch should be described by 
the fully noncommutative phase of $U(N_{\rm BH})$ gauge theory with $R_{{\rm S}^3}=(\lambda_{\rm BH}/\lambda)^{1/4}$;
here we identified the S$^3$ with the surface right outside the bunch.\footnote{

If this rescaling is not performed, then the number of D-branes in the bunch (i.e. $N_{\rm BH}$) can change 
as the energy grows, and hence the identification with the `smallest large black hole' fails. 
}
This scaling of $R_{{\rm S}^3}$ naturally suggests the scalings in \eqref{eq:NBH-dependence}. 
Note that the 't Hooft coupling changes from $\lambda=g_{\rm YM}^2N$ to $\lambda_{\rm BH}=g_{\rm YM}^2N_{\rm BH}$;  
this does not affect the result because the quantities of interest do not explicitly depend on the 't Hooft coupling.

In order to calculate the energy, entropy, and justify \eqref{eq:NBH-dependence} more quantitatively, 
let us appeal to the AdS/CFT duality for large BH from here on, 
and go to the gravity picture.  
{\it If we assume the dual gravity calculation of the large BH to be correct}\footnote{
This is the only assumption which relies on the dual gravity description. 
Note that we assumed here the validity of the dual gravity description for the {\it large} black hole in order to derive (\ref{eq:tadsbh} --\ref{eq:emin}) 
and will derive the energy of the {\it small} black hole in the following. 
}, when $\lambda=g_{\rm YM}^2N\gg 1$,  
the temperature of AdS$_5$ black hole is given by 
\begin{eqnarray}\label{eq:tadsbh}
T_{\rm AdS-BH}=
\frac{2r_+^2+R_{\rm AdS}^2}{2\pi R_{\rm AdS}^2r_+},
\end{eqnarray} 
which is minimized at $2r_+^2=R_{\rm AdS}^2=1$,  
\begin{eqnarray}
T_{\rm min}(N,g_{\rm YM}^2)
=
\frac{\sqrt{2}}{\pi R_{\rm AdS}}. 
\label{eq:Tmin}
\end{eqnarray}
The area of the horizon is $2\pi^2 r_+^3$, and hence the entropy is $S_{\rm AdS-BH}=\pi^2r_+^3/(2G_{\rm 5,N})$. 
Here $G_{\rm 5,N}$ is the 5d Newton constant, which is related to the 10d Newton constant\footnote{
This is different from usual value in the Einstein frame, $G_{\rm 10,N}\sim\lambda^2/N^2$, 
because we are using the dual frame. 
} $G_{\rm 10,N}\sim 1/N^2$, by 
$G_{\rm 10,N}=G_{\rm 5,N}\cdot(\pi^3R_{\rm AdS}^5)$, where the denominator $\pi^3R_{\rm AdS}^5$ is the area of the S$^5$.  
Hence
\begin{eqnarray}\label{eq:s}
S
=
\frac{r_+^3\pi^5R_{\rm AdS}^5}{2G_{\rm 10,N}}. 
\end{eqnarray}
By using $dE=TdS$, we obtain 
\begin{eqnarray}\label{eq:e}
E
=
\frac{3\pi^4 R_{\rm AdS}^3}{8G_{\rm 10,N}}(r_+^4+R_{\rm AdS}^2r_+^2). 
\end{eqnarray}
$E_{\rm min}(N,g_{\rm YM}^2)$ and $S_{\rm min}(N,g_{\rm YM}^2)$ are given by using $r_+^2\sim R_{\rm AdS}^2=1$  and we obtain
\begin{eqnarray}\label{eq:emin}
E_{\rm min}(N,g_{\rm YM}^2)
\sim
\frac{R_{\rm AdS}^7}{G_{\rm 10,N}},
\qquad
S_{\rm min}(N,g_{\rm YM}^2)
\sim
\frac{R_{\rm AdS}^8}{G_{\rm 10,N}}. 
\label{eq:Emin}
\end{eqnarray}
When the bunch shrinks, $R_{\rm AdS}=1$ should be replaced by\footnote{
Because the curvature radius changes, 
this is different from the `truncation' of the geometry. 
} $R_{\rm AdS}'=(N_{\rm BH}/N)^{1/4}$ associated with the rescaling of $R_{\rm S^{3}}$. The Newton constant remains unchanged, because the rescaling factors associated with $R_{\rm AdS}\to R_{\rm AdS}'$ 
and $N\to N_{\rm BH}$ cancel with each other.  
Hence we should have \eqref{eq:NBH-dependence} again.\footnote{
This kind of matching has a flavor similar to the correspondence principle \cite{Susskind:1993ws,Horowitz:1996nw}.  
}

We identify the energy and entropy of the small black hole with these values: 
\begin{eqnarray}
E_{\rm BH}=E_{\rm min}(N_{\rm BH},g_{\rm YM}^2), 
\qquad
S_{\rm BH}=S_{\rm min}(N_{\rm BH},g_{\rm YM}^2). 
\end{eqnarray}
Then,  
\begin{eqnarray}
T_{\rm BH}
=
\frac{dE_{\rm BH}}{dS_{\rm BH}}
\sim
\alpha^{-1/4}. 
\label{temp_BH}
\end{eqnarray}
Note that this might be different from $T_{\rm min}(N_{\rm BH},g_{\rm YM}^2)$. 
By substituting \eqref{temp_BH} into \eqref{eq:NBH-dependence}, we obtain
\begin{eqnarray}
E_{\rm BH}
\sim
\frac{N^2}{T_{\rm BH}^7}
\sim
\frac{1}{ G_{\rm 10,N}T_{\rm BH}^7}. 
\end{eqnarray}

Before closing this section, let us give a comment on a confusing point associated with the evaluation of \eqref{eq:NBH-dependence} via the U($N_{\rm BH}$) theory.  
When $N_{\rm BH}$ decreases, if one naively `truncated' the theory to the U($N_{\rm BH}$) theory without rescaling $R_{{\rm S}^3}$, 
one would not have an $N_{\rm BH}$ dependence. 
This treatment is wrong because the truncation and the variation of the energy do not commute.
When  energy is added, in the original theory
$N_{\rm BH}$ increases and as a result the temperature can go down,
while in the truncated theory $N_{\rm BH}$ cannot change and the temperature has to go up.
The argument on the scaling of eigenvalues, which is provided at the beginning of this section, 
may seem to suffer from the same subtlety. However, we do not find a problem  
there, because we did not change the energy, rather we varied $g_{\rm YM}^2$.  
The result \eqref{eq:NBH-dependence} does not explicitly depend on $g_{\rm YM}^2$; 
namely the bunch size is determined solely by the energy.

\subsubsection{The phase transition from the big black hole to the small black hole}
\hspace{0.25in}
According to the calculation in the gravity side \cite{Dias:2016eto}, the transition from the large black hole to the small black hole
is of first order, and the small black hole is hotter than the large black hole at the same energy. (More precisely speaking, the large BH becomes 
a lumpy black hole \cite{Dias:2015pda}, and then becomes unstable.)

In our gauge theory argument, 
the order of the transition is not clear. However, if we assume the transition is of first order, then the small black hole must be hotter,
due to the Higgsing.

\subsection{The Hagedorn growth at $N_{\rm BH}/N \lesssim 1/\lambda$ ($\lambda\gg 1$)}\label{sec:Hagedorn}
\hspace{0.25in}
In the previous section, we have assumed $\lambda_{\rm BH}\equiv g_{\rm YM}^2N_{\rm BH}\gg 1$. 
When $\lambda_{\rm BH}\lesssim 1$, we can use perturbation theory. 
There, $T_{\rm min}$, which is of same order as the deconfining and hagedorn temperatures, is of 
O($N^0$)
and goes to a $\lambda_{\rm BH}$-independent constant. 
The growth of the temperature stops when $N_{\rm BH}/N\sim 1/\lambda$, at $T_{\rm BH}\sim 1$, $E/N^2\sim 1/\lambda^2$. 
The constant-$T$ behavior below this point looks like the Hagedorn behavior; 
actually, in the closed string picture, the length of the long string $N_{\rm BH}^2$ increases with $E$. 
This is exactly the Hagedorn behavior! 
The energy $E$ and the entropy $S(E)$ are proportional to each other, $E_{\rm BH}\sim S_{\rm BH}\sim N_{\rm BH}^2$. 
In other words, when energy is added, it is used for exciting more matrix degrees of freedom, 
rather than increasing the energy per degree of freedom. 

Note also that, at $N_{\rm BH}/N \sim 1/\lambda$, the energy becomes $E\sim 1/(l_{\rm s}^7G_{\rm 10,N})$, 
which is the endpoint of the Hagedorn growth expected on the gravity side \cite{Aharony:1999ti}. 

\subsection{The case of weak coupling $\lambda\ll 1$}
\begin{figure}[htbp]
\begin{center}
\rotatebox{0}{
\scalebox{0.6}{
\includegraphics{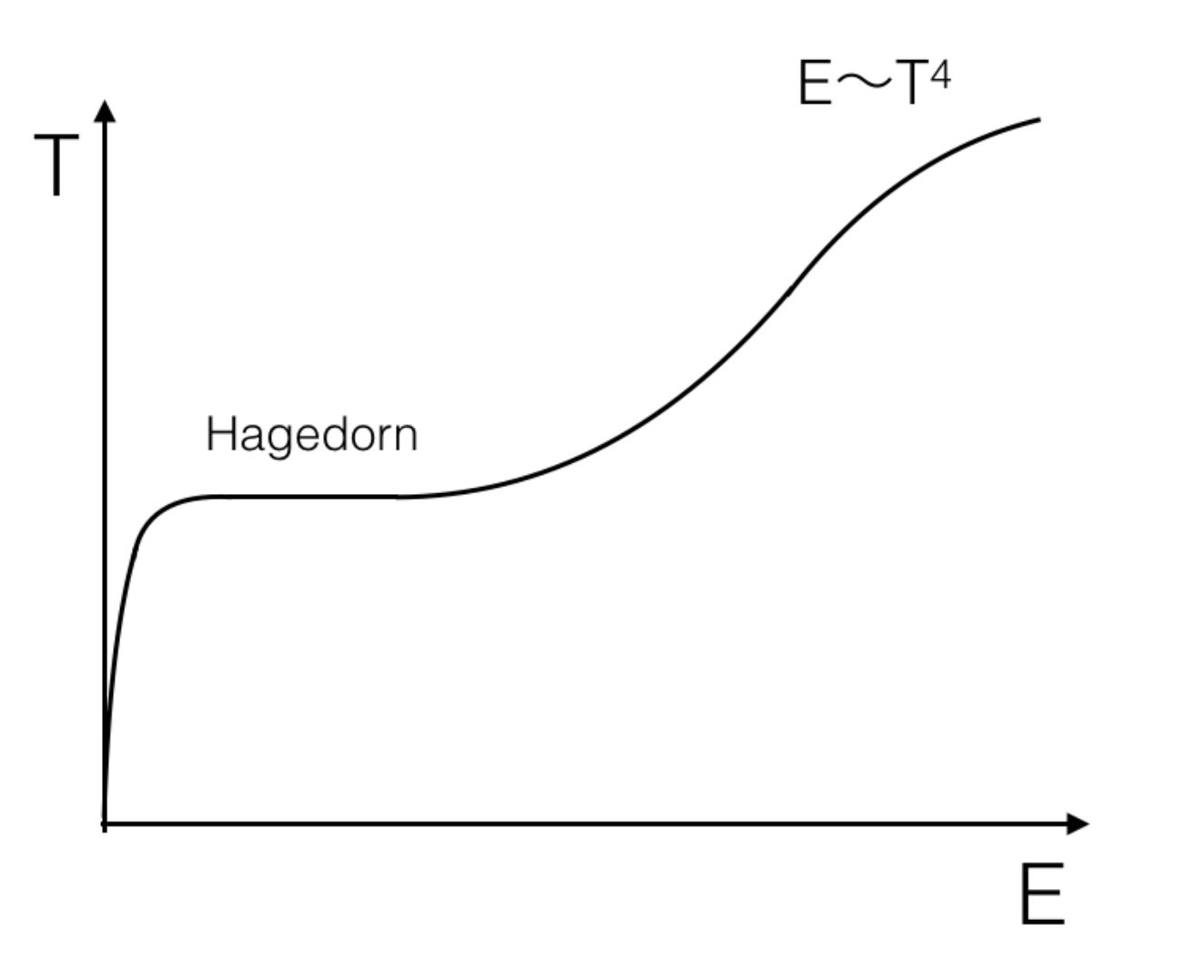}}}
\end{center}
\caption{Phase diagram of the weakly coupled 4d SYM, $\lambda\ll 1$. 
The difference from the one at $\lambda\gg 1$ (Fig.~\ref{phase_diagram}) is not just an overall factor.}\label{phase_diagram_weak_coupling}
\end{figure}

\hspace{0.25in}The same idea of emission of eigenvalues can be applied to the weakly coupled region of 4d SYM. 
In this region the Hagedorn growth continues until $N_{\rm BH}$ reaches $N$, 
and hence a negative specific heat is not expected;  
see Fig.~\ref{phase_diagram_weak_coupling}. 
This region is rather different from the strong coupling region (Fig.~\ref{phase_diagram}). 
The difference is not just a factor $3/4$ rather the shape is different. 

This phase diagram Fig.~\ref{phase_diagram_weak_coupling} has been known for quite some time, see e.g. \cite{Aharony:2003sx}. 
We have just rephrased the known result, in order to show the consistency of our proposal.

\subsection{SO(6) breaking}
\hspace{0.25in}
The 10d BH is localized on the S$^5$. Hence the SO(6) rotational symmetry should be broken. 
A natural possibility would be that the `smallest possible large BH' is a lumpy BH \cite{Dias:2015pda} 
which breaks SO(6) and the eigenvalue distribution in gauge theory side also breaks SO(6).  
More analysis is desirable concerning this point.

\section{Other cases}\label{sec:others}

\subsection{ABJM theory and 11d black hole}
\hspace{0.25in}The argument shown above can be applied to other quantum field theories as well. 
As an example, let us consider the M-theory region of the ABJM theory \cite{Aharony:2008ug} on $\mathbb{R}^1\times$S$^2$.
Namely we consider the Chern-Simons level $k=1$, 
't Hooft coupling $\lambda=N/k=N$. 
The gravity dual is M-theory on AdS$_4\times$S$^7$.  
In this case, the microscopic picture on the gravity side is not clear, other than that $N$ corresponds to the number of M2-branes. 
However it would be natural to {\it assume} that the small BH is described in the same way as in 4d ${\cal N}=4$ SYM, 
by a bunch of $N_{\rm BH}$ M2-branes, and the calculation goes through in the same manner. 
 
With this assumption, by using $R_{\rm AdS}\sim (kN)^{1/6}l_{\rm P}=N^{1/6}l_{\rm P}$ and 
$G_{11,{\rm N}}\sim l_{\rm P}^{9}$ where $G_{11,{\rm N}}$ and $l_{\rm P}^{9}$ are the eleven-dimensional 
Newton constant and Planck scale, respectively, the energy and entropy are estimated as\footnote{
For 4d ${\cal N}=4$ SYM, we used the fact that the eigenvalues scales as $\lambda^{1/4}$. 
Here, we need to use a similar relation: 
the bifundamental scalars $\phi$, which describes the moduli space of M2-branes, should scale as $N^{1/6}$. 
In the 't Hooft limit ($N/k=O(N^0)$), 
because $\phi$ has a potential of the form $N\phi^6/\lambda$, the scaling should be $\lambda^{1/6}=(N/k)^{1/6}$. 
The same behavior is expected in the M-theory region, up to $1/N$-suppressed corrections \cite{Azeyanagi:2012xj}. 
} 
\begin{eqnarray}
E_{\rm min}
\sim
\frac{R_{\rm AdS}}{G_{4,{\rm N}}}
\sim
\frac{R_{\rm AdS}^8}{G_{11,{\rm N}}}
\sim
\frac{N^{4/3}}{l_{\rm P}}, 
\qquad
E_{\rm BH}\sim \frac{N_{\rm BH}^{4/3}}{l_{\rm P}}. 
\end{eqnarray}
and 
\begin{eqnarray}
S_{\rm min}
\sim
\frac{R_{\rm AdS}^2}{G_{4,{\rm N}}}
\sim
\frac{R_{\rm AdS}^9}{G_{11,{\rm N}}}
\sim N^{3/2},  
\qquad
S_{\rm BH}\sim N_{\rm BH}^{3/2}. 
\end{eqnarray}
Therefore, 
\begin{eqnarray}
T_{\rm BH}
=
\frac{dE_{\rm BH}}{dS_{\rm BH}}
\sim 
N_{\rm BH}^{-1/6}/l_{\rm P}, 
\end{eqnarray}
yielding finally
\begin{eqnarray}
E_{\rm BH}
\sim
\frac{1}{G_{11,{\rm N}}T_{\rm BH}^8}. 
\end{eqnarray}
This correctly reproduces the property of the eleven-dimensional Schwarzschild black hole.
In this case (ABJM, $k=1$) the system is strongly coupled even for small $N_{\rm BH}$, 
and hence the Hagedorn behavior discussed in Sec.~\ref{sec:Hagedorn} does not set in. 
\subsection{More generic theories}
\hspace{0.25in}The same power counting will hold for other theories as well, 
including non-conformal theories, if they possess a similar matrix description. 
We assume that the geometry consists of $d$ non-compact and $D-d$ compact dimensions 
(where $D=10$ and $D=11$ for theories with respectively string and M-theory duals). 
The space-time does not necessarily have to be a product like AdS$\times$X (where X is a compact manifold), 
as long as the notion of large and small black holes still makes sense.
Furthermore, suppose there is only one typical length scale $R$ in the dual geometry, like $R_{\rm AdS}$.  
Then, it is natural to expect that the minimum energy and temperature are obtained by the dimensional analysis as 
\begin{eqnarray}
E_{\rm min}\sim\frac{R^{D-3}}{G_{D,{\rm N}}}, 
\qquad
S_{\rm min}\sim\frac{R^{D-2}}{G_{D,{\rm N}}}. 
\end{eqnarray}
By going to the dual field theory description and applying the same argument for the small black hole, we obtain 
\begin{eqnarray}
E_{\rm BH}\sim\frac{(R\alpha)^{D-3}}{G_{D,{\rm N}}}, 
\qquad
S_{\rm BH}\sim\frac{(R\alpha)^{D-2}}{G_{D,{\rm N}}},  
\end{eqnarray}
and 
\begin{eqnarray}
T_{\rm BH}\sim 1/(R\alpha), 
\end{eqnarray}
where $\alpha=(N_{\rm BH}/N)^p$. The power $p$ may depend on the theory.  
This gives the expected scaling, 
\begin{eqnarray}
E_{\rm BH}\sim\frac{1}{G_{D,{\rm N}}T_{\rm BH}^{D-3}}, 
\qquad
S_{\rm BH}\sim\frac{1}{G_{D,{\rm N}}T_{\rm BH}^{D-2}}. 
\end{eqnarray}
%
\section{Discussions}
\hspace{0.25in}A rather striking consequence of our proposal is that the small BH is essentially like a large BH from the point of view of the gauge theory; it is the `smallest possible large BH', 
which is continuously connected to the high-$T$ region. 
Seen from gauge theory, it is simply a thermal state, but with different `matrix size'. 
Hence the study of the large BH provides us with important lessons on the small BH. 
Another lesson is the importance of eigenvalue dynamics in the gauge/gravity duality a la Maldacena. 
Although the importance of eigenvalue dynamics has been widely appreciated in the 20th century, 
for example in the Matrix Theory conjecture \cite{Banks:1996vh}, 
it has somehow almost been forgotten after the Maldacena conjecture (AdS/CFT). 
It is, however, an important piece for understanding black hole evaporation,  
even in the context of the Maldacena conjecture, as emphasized in this paper and Refs.~\cite{Berkowitz:2016znt,Berkowitz:2016muc}. 
Due to this, a detailed study of eigenvalue dynamics should help lead us to an understanding of the bulk geometry, 
including the horizon of the black hole. 
Note that the large BH can be studied by using the canonical ensemble, which makes the numerical simulation 
rather straightforward with the Matsubara formalism. 
It would provide us with a first-principle study of the geometric structure of the Schwarzschild black hole 
based on gauge theory \cite{horizon}. Such a study should be important for various problems associated with the black hole information puzzle. 

More tests would be desirable to establish the proposal more rigidly. 
One interesting and doable direction would be a consistency check based on dual gravity calculations. 
If our proposal is correct, the small black hole and the large black hole at `$E_{\rm min}$' describe essentially the same 
dual gauge theory, up to the rescaling of the 't Hooft coupling associated with the emission of D-branes.  
Recently, dual gravity calculations for the small black hole have been performed in \cite{Dias:2016eto}. 
There is also an attempt for studying the small AdS$_5$ black hole, which is not localized along S$^5$ which can be found in 
\cite{Jokela:2015sza}.
Since the agreement should become better when 10d black hole is smaller, such tests might be doable without 
relying on the very hard numerical calculations needed to include the finite-size effects \cite{Dias:2016eto}. 
We have not yet understood how the breakdown of SO(6) symmetry can be seen in terms of eigenvalues. 
It would be nice if we could make progress in near future. 

In this paper we truncated large matrices to small matrices. More refined treatments, for example something like 
the matrix renormalization group \cite{Brezin:1992yc} which integrates out the emitted eigenvalues, 
would allow us to extract more information about black holes.

\begin{center}
{\bf Acknowledgments}
\end{center}
\hspace{0.25in}
We would like to thank Leonard Susskind for the collaboration toward the end of the project. 
As explained in the introduction, he had essentially the same idea as ours.  
We would also like to thank Juan Maldacena and Kostas Skenderis for pointing out miscalculations
in the first version of this paper, which helped us correcting a few wrong arguments. 
Furthernore we thank Ahmed Almheiri, Tom Banks, Evan Berkowitz, Oscar Dias, Blaise Gout\'{e}raux, 
Guy Gur-Ari, Joao Penedones, Enrico Rinaldi, Jorge Santos, 
Andreas Sch\"{a}fer, Stephen Shenker, David Tong and Ying Zhao for discussions and comments.
The work of M.~H. is supported in part by the Grant-in-Aid of the Japanese Ministry 
of Education, Sciences and Technology, Sports and Culture (MEXT) 
for Scientific Research (No. 25287046). 
The work of J.~M. is supported by the California Alliance fellowship (NSF grant 32540).


\end{document}